# Characterizing Earth-like planets with Terrestrial Planet Finder

S. Seager[a]

[a]Department of Terrestrial Magnetism, Carnegie Institution of Washington, Washington, DC 20015

E. B. Ford and E. L. Turner[b]

[b]Princeton University Observatory, Princeton, NJ 08544

## ABSTRACT

For the first time in human history the possibility of detecting and studying Earth-like planets is on the horizon. Terrestrial Planet Finder (TPF), with a launch date in the 2015 timeframe, is being planned by NASA to find and characterize planets in the habitable zones of nearby stars. The mission Darwin from ESA has similar goals. The motivation for both of these space missions is the detection and spectroscopic characterization of extrasolar terrestrial planet atmospheres. Of special interest are atmospheric biomarkers—such as $O_2$, $O_3$, $H_2O$, CO and $CH_4$—which are either indicative of life as we know it, essential to life, or can provide clues to a planet's habitability. A mission capable of measuring these spectral features would also obtain sufficient signal-to-noise to characterize other terrestrial planet properties. For example, physical characteristics such as temperature and planetary radius can be constrained from low-resolution spectra. In addition, planet characteristics such as weather, rotation rate, presence of large oceans or surface ice, and existence of seasons could be derived from photometric measurements of the planet's variability. We will review the potential to characterize terrestrial planets beyond their spectral signatures. We will also discuss the possibility to detect strong surface biomarkers—such as Earth's vegetation red edge near 700 nm—that are different from any known atomic, molecular, or mineralogical signature.

## 1. INTRODUCTION

The field of extrasolar planets has grown tremendously over the past several years with over one hundred extrasolar giant planets now known to orbit nearby sun-like stars. Current technology can only detect giant planets around sun-like stars but ambitious space missions TPF and Darwin are being designed to not only detect terrestrial-like planets but also to study their atmospheres via low-resolution spectroscopy.

Extrasolar planets are one of the few astronomical objects with local counterparts—the solar system planets. Using the solar system planets and the physics and computational algorithms developed to understand them we are predicting extrasolar planet atmosphere signatures and preparing to interpret atmospheric data. One example of this is the first—and only to date—extrasolar planet atmosphere detection of neutral sodium absorption in the transiting extrasolar giant planet HD209458b [1]. This atmosphere detection of the transiting planet is reassuring because it involved a specialized observation that was possible only with the advance knowledge of what atmospheric feature to look for (as first shown in [2]). The success of this detection shows that applications of atmospheric physics to planets in new environments can be successful and that their application to experimental design is reliable.

Terrestrial-like extrasolar planet atmosphere signatures may be far more difficult to predict and also to interpret than extrasolar giant planet atmosphere signatures. This is because the giant planet atmospheres are primitive—little atmospheric evolution has taken place so that they contain roughly the same atmospheric gases as at their formation. Indeed Jupiter and Saturn (and to some extent the ice giants Uranus and Neptune) have similar compositions and even to first order similar spectral signatures. In contrast the four terrestrial-like solar system planet atmospheres are all significantly different, due to atmospheric evolution. Both atmospheric escape of light gases and gas-surface reactions have substantially affected each of the terrestrial planet atmospheres. Thus we expect terrestrial-type planets to be much more diverse and extremely interesting for study, in comparison to the extrasolar giant planets. While the primary TPF goal is to detect terrestrial-like planets and to spectroscopically characterize them, much more physical characterization is possible with some work and some flexibility in the data acquisition.

## 2. BIOMARKER GASES

The most exciting outcome of a TPF search for extrasolar planets would be a spectroscopic detection of an atmosphere in severe chemical nonequilibrium, for example the simultaneous presence of reducing and oxidizing gases [3] or of species that are compatible with life. Earth is our only example of planet with an atmosphere modified by life so, following [4], we use it as a basis for discussion. On Earth both $O_2$ and its photolytic product $O_3$ have strong spectral features. $O_2$ is highly reactive and remains in the Earth's atmosphere only because it is continually produced by vegetation respiration. No abiological sources of continuous, large quantities of $O_2$ are known. Thus $O_2$ and $O_3$ are considered to be the best biomarker gases, albeit for life as we know it. Any highly reactive species, however, detected in an extrasolar planet atmosphere in great abundance will be of huge interest. $N_2O$ is a second gas produced by life during microbial oxidation-reduction reactions, but is produced in small quantities and has only a very weak spectroscopic signature that overlaps with $H_2O$ and $CH_4$ absorption bands. On Earth $H_2O$ vapor is abundant and is readily apparent in Earth's atmosphere (this is well-known by ground-based astronomers). Because all life as we know it requires liquid water, detection of $H_2O$ vapor is very important and is considered to be the second most reliable biomarker gas. Note that $H_2O$ is not necessarily indicative of life but is compatible with habitable conditions. Other spectral features, while not biomarkers or habitability indicators, could provide useful information about the planet's character. $CO_2$ should be indicative of a terrestrial-like planet; Earth, Venus, and Mars all have a very strong, saturated $CO_2$ line in the mid-IR. High concentrations of $CH_4$ could indicate the presence of methanogenic bacteria, but could also result from surface processes related to volcanism. The simultaneous detection of combinations of unusual, non-chemical-equilibrium species may be a more robust indicator of life or habitability than detection of only a single unusual feature (e.g., [5]).

## 3. RADIUS AND MASS

TPF has the goal to detect extrasolar Earth-like planets and to characterize them spectroscopically. Planetary radii and masses are important to know because they provide clues to the planet's formation history, its atmosphere-retaining strength, and its ability to foster life. For example, larger planets sustain higher levels of tectonic activity that also persists for a longer time. Tectonic activity sustains volcanism and also heats crustal rocks and recycles $CO_2$ and other gases back into the atmosphere. These outgassing processes are required to ensure climate stability over geologic timescales [4]. There has been some concern that because TPF will detect flux and not gravitational effects of the planet on the parent star, the mass will not be well characterized. Here we summarize arguments to the contrary, that both planetary radius and mass can be sufficiently well estimated for extrasolar planet science.

At mid-IR wavelengths the planet radius can be estimated by equating the color temperature $T_c$ with the effective temperature ($T_{eff}$) and using the flux measured at Earth:

$$F_p = \int_\Omega \cos\mathbf{q} B_\mathbf{n}(T) d\mathbf{n} d\Omega = \mathbf{s} T_{eff}^4 \approx \mathbf{s} T_c^4,$$

$$F_E = F_p \left(\frac{R_p}{D}\right)^2.$$

Here $B_v$ is blackbody radiation at temperature $T$ and frequency $v$, $F_p$ is the flux at the planet surface, $F_E$ is the flux measured at Earth, and $D$ is the planet-Earth distance. The only unknown in the second equation is hence $R_p$. $T_c$ is derived from a fit to blackbody radiation and is possible at mid-IR wavelengths where the peak of an Earth-like extrasolar planet flux is expected. Note that differences from a blackbody will cause some uncertainty with this estimate.

At visible wavelengths the only flux is due to reflected starlight—the planet has none of its own visible-wavelength emission and hence this wavelength region is far away from the blackbody peak. Nevertheless the planet's size can be estimated from the flux. The main point to note is that gas giant and terrestrial-sized planets can be easily distinguished by their apparent brightness (a function of area, albedo, and phase function) and orbital distance. Considering the surface area ratio of Jupiter to Earth and assuming the same albedo, Jupiter would be 120 times brighter than the Earth at the same orbital distance. Unless giant planet albedos are ~10s to 100 times smaller than terrestrial planet albedos, confusion between giant and terrestrial-size planets is unlikely. More specifically the planet radius can be estimated since the total amount of reflected flux is a combination of the planet area, the albedo ($p$) and the planet phase function $j(a)$,

$$\frac{F_p}{F_*} = p\,j(a)\left[\frac{R_p}{a}\right]^2,$$

where $a$ is the planet-star semi-major axis, $F_*$ is the stellar flux, and $a$ is the phase angle (the planet-star-observer angle). The planet phase function (also sometimes called the phase curve) is a function of planet phase and atmospheric scattering properties (see, e.g. [6]) and can be complex to compute. For example, a Lambert sphere at quadrature ("half full" phase) is 1/3 as bright as a Lambert sphere at opposition (full phase), not 1/2 as bright. The planetary phase angle can be determined from a series of measurements as the planet orbits the star.

The estimated radius can be used together with an assumed density to estimate the planetary mass. The solar system planets generally have a relationship between mass, radius and thermal environment. For example, newly discovered minor planets in our own solar system are assigned radius and mass estimates based on measured photometry and albedo and density assumptions. With this in mind we can estimate an error on the "photometric" mass estimate. For terrestrial planets, the visual geometric albedo varies by roughly a factor of 2.6 from the mean (i.e., a maximum range of 0.09 to 0.6; this covers the variations seen in the geometric albedo in our solar system but ignores differences at quadrature due to the different phase functions (see Figure 4)). A multiplicative albedo error $a_{error}$ leads to a factor $(a_{error})^{1.5}$ on the derived volume and the observed density for terrestrial planets varies by a factor of 1.2 from the mean (i.e., 3.9 to 5.5). The derived mass will therefore have an error of 5.0 and probably much less. A similar error estimate could be obtained for the mid-IR color mass.

There is some concern in the astronomical community that TPF will not be able to measure the planetary mass. Perhaps this concern has arisen from the fact that the minimum mass is currently the only physical parameter known for all but one of the known extrasolar planets. In any event we should remember that the current radial velocity searches only measure the minimum mass and even SIM will not be able to measure mass below five to ten Earth masses. More relevant, planetary migration and formation models require as constraints not precise planetary masses, but a distribution of planet masses with orbital distance. (Note that precise masses are required for dynamical studies of specific systems). Finally it should be realized that mass and radius are only two of the planet's physical properties we want to know and that other properties are arguably more important for habitability and life.

## 4. TEMPERATURE

Temperature is a fundamental characteristic of the planetary atmosphere environment. The temperature determines the phase state of matter (solid, liquid or gas) and governs which species will be present in the atmosphere via equilibrium chemistry or non-equilibrium chemical reaction rates. The temperature is determined from the spectral continuum or lines at mid-IR wavelengths and is determined for the atmospheric region where the continuum radiation originates—at optical depth of unity which can be quite high above the planetary surface. At visible wavelengths the temperature cannot be determined directly and must be inferred from spectral lines (this may be difficult at low resolution).

For extrasolar terrestrial-like planets we would like to determine the surface temperature. This is only possible, however, at mid-IR wavelengths where the atmosphere is transparent. Earth has such a window at 8-12 microns, but a warmer planet (by ~20 K) would not have this spectral window because of continuum absorption by water vapor [4]. Surface properties may be inferred from atmospheric composition and other factors, but for planets with 100% cloud cover, like Venus, the surface temperature would be very difficult to infer. In fact Venus and Earth have very similar effective temperatures (220 K and 255 K respectively) but extremely different surface temperatures (730 K and about 290 K respectively) especially from the perspective of compatibility with life.

## 5. WEATHER, ROTATION RATE, SEASONS, CONTINENTS

A time series of photometric data could reveal a wealth of physical characteristics at wavelengths that penetrate to the planetary surface. Visible wavelengths are likely more suited for these measurements because the albedo contrast of surface components is much greater than the temperature variation across the planet's surface. Moreover, the narrow transparent spectral window at 8-12 microns will close for warmer planets than Earth and for planets with more water vapor than Earth. However, further study at the mid-IR "window" needs to be investigated. See [7] for the details of the model used in this section.

Earth is the most variable planet in our solar system, photometrically speaking. This variability is due to weather: cloud formation, motion, and dissipation. The variability at visible wavelengths is approximately 10 to 20%; and is caused by the contrast in albedo between the reflective clouds and the underlying, darker land or ocean. Evidence of variable water clouds combined with water vapor in the atmosphere is indicative of large bodies of liquid water. A variable planet would definitely warrant further study—compare the variable Earth to the constant, 100% cloud-covered Venus which shows no variability at visible or mid-IR wavelengths.

The rotation rate of a planet is an important physical characteristic because it is a fundamental driver of atmospheric circulation patterns and weather and it is a record of the planet's formation history. The rotation rate can be determined at visible wavelengths on a relatively cloud-free Earth-like planet if the planet has different surface components. The light scattered by such a planet will vary in intensity as the planet rotates, with a repetitive pattern. For example, Earth's major surface components land, ocean, and ice have very different albedos (< 10% for ocean, > 30-40% for land, > 60% for snow and some types of ice), and in the case of a cloud-free Earth viewed at the equator the rotational surface variation can be up to 200% (Figure 1). Even considering the Earth with its cloud patterns, the rotational period is still determinable because large-scale cloud formations persist coherently in some regions for several days (Figure 2).

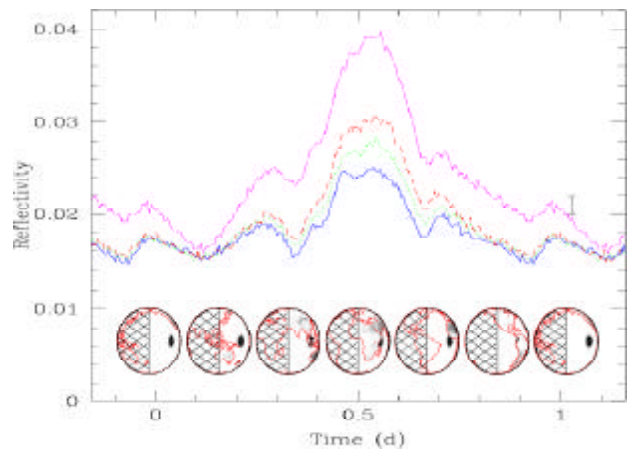

Figure 1. A light curve for a cloud-free Earth model for one rotation. The *x*-axis is time and the *y*-axis is the reflectivity normalized to a Lambert disk at a phase angle of 0°. The viewing geometry is shown by the Earth symbols, and a phase angle of 90° is used. Note that a different phase angle will affect the reflectivity due to a larger or smaller fraction of the disk being illuminated; because of the normalization the total reflectivity is << 1 in this case of phase angle of 90°. From top to bottom the curves correspond to wavelengths of 750, 650, 550, and 450 nm, and their differences reflect the wavelength-dependent albedo of different surface components. The noise in the light curve is due to Monte Carlo statistics in our calculations. The images below the light curve show the viewing geometry (cross-hatched region is not illuminated) and relative contributions from different parts of the disk (shading ranges from < 3% to > 40%, from white to black) superimposed on a map of the Earth. At t = 0.5 day, the Sahara desert is in view and causes a large peak in the light curve due to the reflectivity of sand which is especially high in the near-IR (top curve).

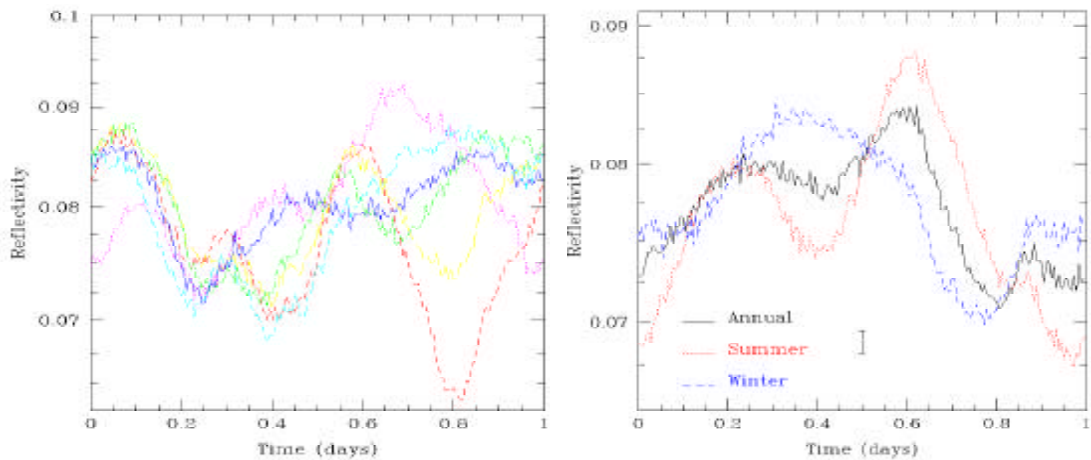

Figure 2 (left) Rotational light curves for model Earth with clouds. This figure shows six different daily light curves at 550 nm for our Earth model with clouds, as viewed from a phase angle of 90 degrees. These theoretical light curves use cloud cover data from satellite measurements taken on six consecutive days. Figure 3 (right) shows theoretical light curves for Earth using seasonal (dotted and dashed lines) and annual (solid line) average cloud cover (averaged over 8 years). Using actual cloud data allows us to accurately model the Earth, but is not applicable to extrasolar terrestrial planets.

Seasonal changes may be difficult to detect from the spectra alone, but with enough photometric data may be extractable from a time series of data. If the planet's rotation rate is known, seasonal changes in the diurnal rotation pattern could be detected by folding the data over rotational periods. For our own Earth, at visible wavelengths the changes may be due to changing ice and snow cover or to changing cloud patterns; both such changes would be indicative of a non-zero planet obliquity. For example, on Earth similar cloud patterns arise during each winter and each summer due to the solar heating angle (Figure 3). At mid-IR wavelengths the changes in the spectral continuum caused by global, seasonal temperature changes are at the 5% level or less [8].

The existence of different surface features on a planet may be discernable at visible wavelengths as long as the cloud cover fraction is not too high. For example, as previously described, considering a cloud-free Earth the diurnal flux variation caused by different surface features rotating in and out of view could be as high as 200% (Figure 1). This high flux variation is not only due to the high contrast in different surface components' albedos, but also to the fact that a relatively small part of the visible hemisphere dominates the total flux from a spatially unresolved planet. Clouds interfere with surface visibility and in the presence of clouds the diurnal light curve shown in Figure 1 becomes that shown in Figure 2. It is very interesting to note that an extrasolar Earth-like planet certainly could have a lower cloud cover fraction than Earth's 50% cloud cover. The cloud pattern and cover fraction are influenced by a variety of factors including the planet's rotation rate, continental arrangement, obliquity, and presence of large bodies of water.

The photometric light curve over the course of a planet's orbital period could reveal information about the atmospheric particles responsible for light scattering. Earth's orbital "phase curve"—taken by observing the reflected Earthlight off of the dark side of the moon [9]—shows that Earth does not scatter light like a Lambert sphere (like most solids) (Figure 4). This is important because the orbital phase curve may constrain the surface or scattering properties in cases where spectra or non-repeated photometric measurements cannot tell the difference between a 100% cloud covered planet like Venus and a completely ice-covered planet.

Simulating time series of short- and long-term photometric variation of an imaginary Earth-like planet— one with a specified continental arrangement, cloud cover fraction and pattern, etc.—is relatively straightforward. The inverse problem of taking a multi-wavelength, long time series of photometric data and deriving the planet surface characteristics is very difficult and may be degenerate in many different parameters (including viewing angle, continental arrangement, ice vs. cloud cover, and more). Much more work is needed to determine whether or not the various degeneracy's can be overcome.

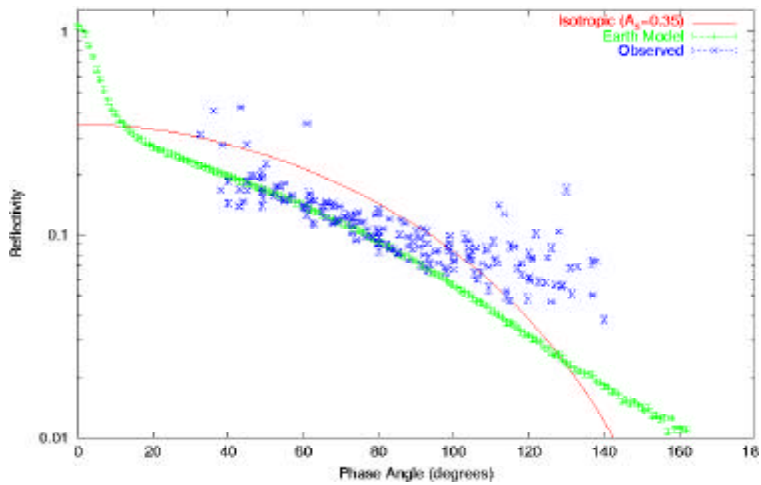

Figure 4. Earth's model orbital phase curve (thick curve) compared to Earthshine data (symbols) from [9]. The scatter in the data increases likely because the measurements are more difficult as the moon approaches full phase. The thin solid line shows a Lambert sphere.

## 6. SURFACE FEATURES AS BIOMARKERS

An extremely exciting possibility is the detection of surface biomarkers in the spectrum of an extrasolar planet. This would be possible at wavelengths that penetrate to the planet's surface, and for surface features, which have large, distinct, abrupt changes in their spectra. Although most surface features (e.g., ice, sand) show very little or very smooth continuous opacity changes with wavelength, Earth has one surface feature with a large and abrupt change: vegetation (Figure 5).

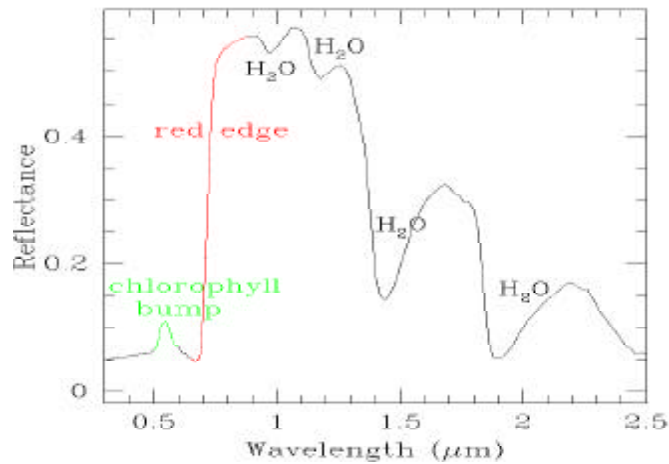

Figure 5—Reflection spectrum of a deciduous leaf. The small bump near 0.5 microns comes from chlorophyll absorption and gives plants their green color. The much larger sharp rise (near 0.8 microns) is known as the "red-edge".

All chlorophyll-producing vegetation has a very strong rise in reflectivity at around 700 nm, a change of almost an order of magnitude! This red-edge spectral signature is much larger than the familiar chlorophyll reflectivity bump at 500 nm, which gives vegetation its green color. Figure 5 shows a deciduous plant reflection spectrum with all of the plant's spectral features. If our eyes could see a little further to the red, the world would be a very different place: plants would be very red, and very bright. The glare from plants would be unbearably high, like that of snow. The red edge is caused both by strong chlorophyll absorption to the blue of 700 nm, and a high reflection to the red of 700nm. The high reflection mechanism allows the plant to not absorb too much energy—overheating will cause chlorophyll to degrade.

The total vegetation red-edge signature is reduced from an order of magnitude reflection spectrum down to a few percent over a spatially unresolved Earth hemisphere. Recent observations of Earthshine to measure the spectrum of the spatially unresolved Earth have detected the red-edge signature at the few percent level [10,11]. This is because of several effects including the canopy, the non-isotropic scattering properties of plants, and the non-continuous coverage of plants across the Earth's surface. A time series of data in different colors (Figure 6) may help make it possible to detect a small but unusual spectral feature. The data would be most useful in the form of a spectrum so that the photometric bands can be chosen after data acquisition.

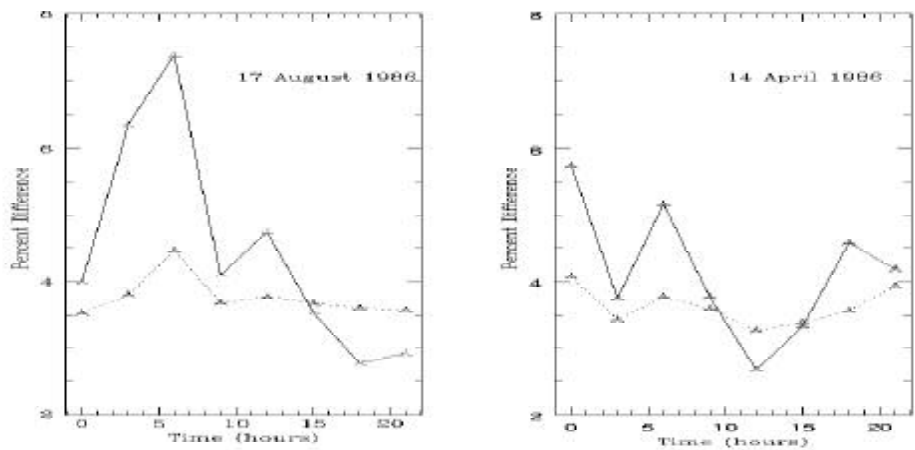

Figure 6. Variability of Earth's color. The solid line shows a color $[(I(0.75-0.8)-I(0.7-0.65))/I(0.75-0.8)]$ chosen to emphasize variability of vegetation's red edge. For comparison, the dotted line shows a color $[I(0.85-0.8)-I(0.75-0.8)/I(0.75-0.8)]$, which is less sensitive to vegetation. These colors include theoretical spectra from [4] modulated by our Earth rotational surface and cloud model. The cloud cover for the model in the left panel is from the ICSSP database from 17 August 1986 and in the right panel from 14 April 1986. This figure shows that Earth is more variable in a color across the red edge than for colors with similar wavelength differences in other parts of Earth's spectrum.

## 7. CONCLUSION

TPF will provide the potential to learn a wealth of information about a terrestrial-like planet's physical properties. The planet radius, mass, and temperature can be constrained. A mission capable of measuring the spectral features of gases important for or caused by life on Earth (e.g. $O_2$, $O_3$, $CO_2$, $CH_4$ and $H_2O$) would have the necessary signal-to-noise necessary to measure photometric variability of the unresolved planet. Such variability could reveal the existence of weather, the planet's rotation rate, presence of different surface features, seasonal changes, and even surface biomarkers.

## 8. ACKNOWLEDGEMENTS

S.S is supported by the Carnegie Institution of Washington. S.S. thanks Nader Haghighipour and Wes Traub for useful discussions.